# A Novel Soil Profile Standardization Technique with XGBoost Framework for Accurate Surface Wave Inversion

### Author 1

- Kousik Mandal, PhD Scholar
- Department of Civil Engineering, Indian Institute of Technology Madras, India
- ORCID number: 0009-0009-5147-3151
- Email: ce22d013@smail.iitm.ac.in

### Author 2

- Tarun Naskar, PhD
- Associate professor at the Department of Civil Engineering, Indian Institute of Technology Madras, India
- ORCID number: 0000-0002-2752-2840
- Email: tarunnaskar@civil.iitm.ac.in

### Contact details of the corresponding author

Tarun Naskar, PhD
Associate professor at the Department of Civil Engineering, Indian Institute of Technology Madras, India.
Email: tarunnaskar@civil.iitm.ac.in

**ABSTRACT**

The inversion of surface wave dispersion curves poses significant challenges due to the non-uniqueness, non-linear, and ill-posed nature of the problem. Traditional local search methods often get trapped in suboptimal minima, leading to lower accuracy, whereas global search methods are computationally intensive, especially when searching large parameter spaces. The increased computational time particularly becomes challenging when dealing with a large number of traces, such as those encountered in 2D/3D surface wave surveys or DAS surveys. Several attempts have been made to perform inversion using machine learning to improve accuracy and reduce computation times. Current machine learning methods rely on a fixed number of soil layers in the training dataset to maintain a consistent output size, limiting these models to predicting only a narrow range of soil profiles. Consequently, no single machine learning model can effectively predict soil profiles with a wide range of shear wave velocities and varying numbers of layers. The present study introduces a novel soil profile standardization technique and proposes a regression-based XGBoost algorithm to efficiently estimate shear wave velocity profiles for stratified media with a varying number of layers. The proposed model is trained using 10 million synthetic soil profiles. This extensive dataset enables our XGBoost model to learn effectively across a wide range of shear wave velocities. Additionally, the study proposes constraints on the differences in shear wave velocities between consecutive layers and on their ratio with layer thickness, preventing the formation of unrealistic layers and ensuring the predictive model reflects real-world conditions. The effectiveness of our proposed algorithm is demonstrated by adopting a wide range of soil profiles from published literature and comparing the results with traditional inversion methods. The model performs well in a wide range of S-wave velocities and can accurately capture any number of layers of the soil profile during the inversion process.

# 1 INTRODUCTION

The Multichannel Analysis of Surface Waves (MASW) method has become a widely adopted geophysical technique for characterizing subsurface properties (Niu et al., 2024; Park et al., 1999). It has been applied extensively across various geotechnical applications, including subsurface exploration, estimation of Vs30, seismic hazard analysis, soil liquefaction assessment, and pavement condition assessment. The MASW method involves three main stages: field data acquisition and extraction of the dispersion curve, forward modelling, and inversion analysis. In the first stage, the wavefield transform technique is employed to generate a dispersion image from the seismogram data collected in the field (Naskar & Kumar, 2022). Free vibration-based forward modelling approaches provide the theoretical modal solution for surface waves propagating through stratified media. The inversion stage then seeks to identify soil profiles whose forward response aligns with the experimental dispersion curve. It employs optimization algorithms that can be categorized into two broad groups: local or global search techniques.

Local search methods rely on the Jacobian matrix of phase velocity to iteratively adjust soil properties in order to match the experimental dispersion curve (Abbiss, 2001; Ganji et al., 1998; Lin et al., 2022; Xia et al., 1999). While computationally efficient, local search methods are highly sensitive to the initially assumed soil properties and frequently get trapped in local minima, resulting in suboptimal solutions (Socco et al., 2010). In contrast, global search methods explore a predefined parameter space to identify the best-fit soil properties (Beaty et al., 2002; Foti, 2003; Le et al., 2024; Sambridge, 1999; Socco & Boiero, 2008). However, global search methods are computationally expensive; therefore, they are sporadically employed by practising engineers in the field. The high computational time becomes especially problematic in 2D or 3D MASW surveys and Distributed Acoustic Sensing (DAS) based surface wave surveys, where dispersion curves must be extracted at hundreds or even thousands of spatial locations. Each location requires a separate inversion, and the cumulative computational load quickly becomes impractical. Nevertheless, experts still prefer global search methods due to their independence from starting soil parameters and ability to avoid local minima.

With recent progress in machine learning (ML), a few attempts have been made to perform inversion using ML to address the low inversion accuracy of local search methods and the long computational time associated with global search approaches. Different ML algorithms have been explored to map the dispersion curve to the layered soil properties. Convolutional neural networks (CNNs) use kernels to automatically extract important features from the dispersion curve and use neural network layers to learn complex nonlinear mapping (Chen et al., 2022; Hu et al., 2020). Mixture density neural networks (MDNNs) have been used to provide a probabilistic distribution of the inverted soil profile rather than a single deterministic output (Cao et al., 2020; Earp et al., 2020). However, these distributions serve as meaningful uncertainty estimates of the target dispersion curve only when validated through the forward model. Furthermore, all these aforementioned studies considered a fixed number of layers with constant thickness values, therefore significantly limiting their usefulness. Yablokov et al., (2021) trained an artificial neural network (ANN) for surface wave inversion that predicts both layer thickness and S-wave velocity. Their approach uses a layering ratio to set thickness ranges, which defines the number of layers for inversion, while velocity ranges are set according to their proposed algorithm. However, the concept of layering ratio is site-specific, which restricts the application of their ANN model across diverse soil profiles. Wu et al., (2022) introduced the First Height Last Velocity (FHLV) custom loss function in a Long Short-Term Memory (LSTM) framework, significantly enhancing the accuracy of the thickness prediction compared to the traditional mean absolute error (MAE) loss function. However, their approach is limited only to a narrow range of profiles. Keil & Wassermann, (2023) proposed a two-step inversion approach to address the challenges associated with accurately predicting both the number of layers and shear wave velocity. First, they employed a classification neural network to predict the number of layers. Based on this prediction, they applied a MDN model trained separately for soil profiles with two to seven layers, allowing simultaneous prediction of layer velocity and thickness. As a result, using Keil & Wassermann, (2023) approach, it is not possible to train a single machine learning model that can handle a wide range of shear-wave velocities while accommodating varying numbers of layers. Moreover, their results indicate that for soil profiles with more than two layers, the half-space velocity is notably high, often exceeding 1500 m/s. Additionally, they imposed a constraint on the first layer, limiting the velocity range to 100-500 m/s. Furthermore, they maintained a minimum layer thickness of 10 m for intermediate layers, which subsequently reduced the thickness variation for the higher layers. Thus, the constraint dataset may lead to apparent higher accuracy for their prediction; however, its performance may be limited in a realistic scenario. In summary, current machine learning models are unable to handle inconsistent output formats. Furthermore, when soil profiles contain an unknown number of layers and wide velocity ranges, they struggle to accurately predict the soil profile. As a result, developing a single ML model that can accurately predict a wide range of shear-wave velocities across varying layer configurations remains a significant challenge.

This paper presents a novel soil profile standardization technique combined with a regression-based XGBoost algorithm to accurately and efficiently predict layer thickness and S-wave velocity for a wide range of layered media. The proposed standardization technique converts soil profiles with varying numbers of layers into a uniform ten-layer format, with predefined layer thicknesses that increase geometrically with depth. It eliminates the need to explicitly estimate the number of layers during inversion while preserving a high level of accuracy and enabling reliable prediction of layer thicknesses. Moreover, this transformation ensures consistent output dimensions for our model, thereby addressing a major limitation of existing ML techniques. The XGBoost was selected for its commendable performance with structured data and its ability to effectively model complex relationships within the dataset (Chen & Guestrin, 2016). The proposed model is trained using 10 million synthetic soil profiles, with shear wave velocities ranging from 100 m/s to 1200 m/s. This extensive dataset provides good parameter space coverage and minimizes sampling bias that prevents the existing model from learning artifacts caused by data sparsity. The Vs30 plays a crucial role in engineering design; therefore, we decided to keep the half-space depth at 30 m, however, it can be changed as per the requirement. The proposed model demonstrated high accuracy on the test data, achieving an r2 score of 0.9788. Inversion results for 14 diverse soil profiles adopted from the literature are compared with the commercial software and the state-of-the-art global search-based tool, Geopsy. The proposed XGBoost model significantly outperforms the existing LSM and is comparable in accuracy to Geopsy. This is a commendable achievement, considering that existing machine learning methods have so far demonstrated good performance only over a limited range of soil profiles close to their training model. Readers are reminded that example profiles are significantly different from the training data set, and a few contain characteristics that the model has never seen. Moreover, our trained XGBoost model requires only a fraction of a second while delivering accuracy that matches or surpasses state-of-the-art GSM. This drastic reduction in computational time while maintaining high accuracy is especially beneficial for 2D/3D MASW and DAS-based surface wave surveys, where dispersion curves must be inverted for numerous traces. The combination of speed and accuracy of the proposed XGBoost model will enable practising engineers to perform rapid, precise inversion across entire survey grids, significantly reducing data processing from weeks to a few hours.

## 2. Methodology

### 2.1 Training data generation

Training data generation process involves standardizing soil profiles to ensure consistent output dimensions while preventing unrealistic soil profile generation. Profiles with varying numbers of layers are converted into a ten-layer format. Each shear wave velocity value is randomly repeated across one or more consecutive layers in the standardized profile while preserving the original layer order (Fig. 1).

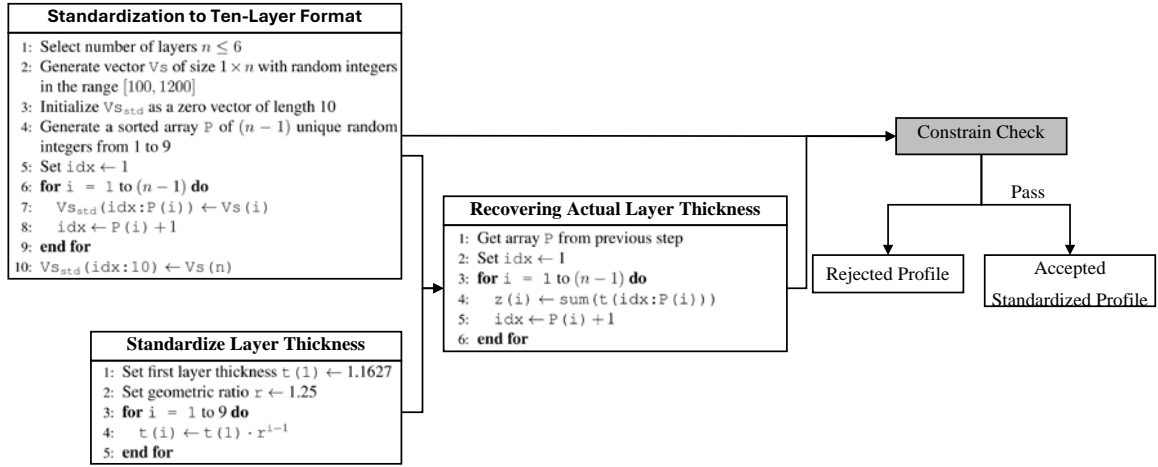

**Fig. 1.** Flowchart of the training data generation process.

All standardized layers except the last are assigned a constant thickness that follows a geometric progression with a layering ratio of 1.25, with the sum of thicknesses equal to 30 m. To ensure realistic soil profiles, the following constraints are imposed:

$$V_{s,i+1} - V_{s,i} \in [-250 \, m/s, 600 \, m/s] \tag{1}$$

$$\frac{|V_{s,i} - V_{s,i-1}|}{z_i} \in [5, 300] \tag{2}$$

where $V_{s,i}$ represents the original shear wave velocity of the $i$-th layer and $z_i$ denotes the recovered thickness of the corresponding layer. Equation 1 limits the allowable difference in shear wave velocity between consecutive layers, avoiding abrupt velocity changes. Equation 2 represents the ratio of the absolute velocity difference between the previous and current layers to the thickness of the current layer. The lower bound prevents small velocity changes over thick layers, and the upper bound prevents large velocity changes over thin layers. Additionally, the maximum allowable velocity reversal is restricted to two, eliminating oscillatory patterns in layered profiles, and the half-space shear wave velocity is kept higher than the overlying layers. In Fig. 2, the standardization process is demonstrated using a four-layered soil profile.

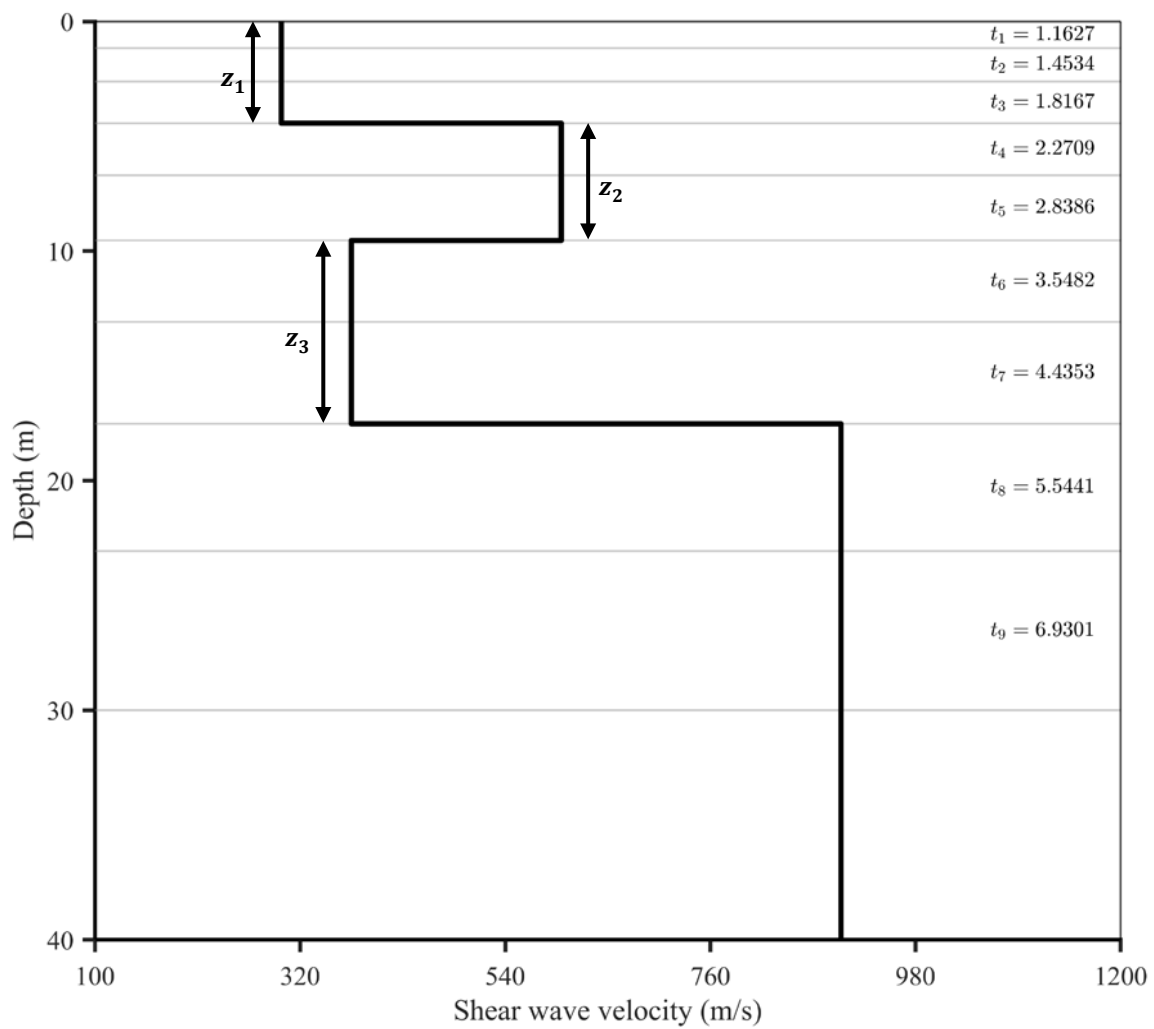

**Fig. 2.** The soil profile standarization technique: Transformation of a four-layer soil profile into a standardized ten-layer format.

Dispersion curves are generated for the accepted profile using the fast delta matrix method (Buchen & Ben-Hador, 1996) over a frequency range of 1 Hz to 100 Hz with a resolution of 1 Hz; however, the model can be trained for any custom frequency range if required. A constant Poisson's ratio of 0.35 and unit weight of 20 $kN/m^3$ are adopted, as these parameters are less sensitive to Rayleigh wave dispersion.

## 2.2 Core XGBoost Algorithm

XGBoost is an advanced implementation of the gradient boosting algorithm, designed for efficient training on big datasets using approximate split-finding methods to minimize a regularized loss function. In Fig. 3, a schematic representation of the XGBoost training and

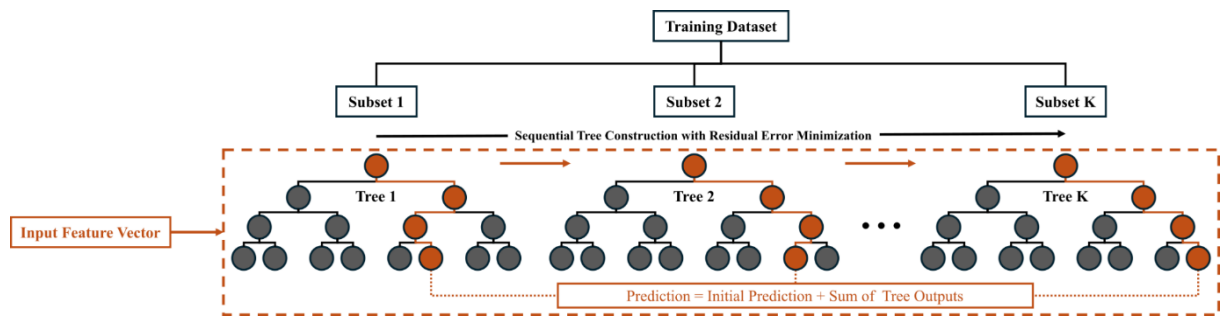

**Fig. 3.** Flowchart illustrating the training and prediction workflow of the XGBoost model.

prediction process is provided. The regularized loss function ($\mathcal{L}$) can be expressed as:

$$\mathcal{L} = \sum_{i=1}^{n} l(y_i, \hat{y}_i) + \sum_{k=1}^{K} \Omega(f_k) \qquad (3)$$

where $y_i$ is the true target value, $\hat{y}_i$ is the predicted value, $i$ represents the current sample number, $n$ is the total number of sample, $l$ is the loss function that measures the error between $y_i$ and $\hat{y}_i$, $f_k$ denotes the $k$-th regression tree, $\Omega(f_k)$ is the regularization term, and $K$ represents the total number of trees. $\Omega(f_k)$ penalizes model complexity to prevent overfitting:

$$\Omega(f) = \gamma T + \frac{1}{2} \lambda \sum_{j=1}^{T} w_j^2 \qquad (4)$$

$T$ is the number of leaves, $w_j$ is the weight of the leaf $j$, and $\gamma$ and $\lambda$ are regularization parameters. XGBoost grows trees in series, adding a new tree $f_t$ at iteration $t$ to predict the residuals from earlier trees. The objective function on iteration $t$ is:

$$\mathcal{L}^t = \sum_{i=1}^{n} l\left(y_i, \hat{y}_i^{(t-1)} + f_t(x_i)\right) + \Omega(f_t) \tag{5}$$

where $\hat{y}_i^{(t-1)}$ being the prediction from the previous trees. The loss function $\mathcal{L}^t$ can be approximated using a second-order Taylor expansion. By setting the derivative of this approximated loss function with respect to the leaf weights to zero, we can find the optimal weights for each leaf as:

$$w_j^* = -\frac{\sum_{i \in I_j} g_i}{\sum_{i \in I_j} h_i + \lambda} \tag{6}$$

where $I_j$ represents indices of instances in leaf $j$. $g_i$ and $h_i$ are gradients and Hessians of the loss function, respectively. After the split, the loss reduction (also known as gain) can be expressed as:

$$Gain = \frac{1}{2}\left(\frac{\left(\sum_{i \in I_L} g_i\right)^2}{\sum_{i \in I_L} h_i + \lambda} + \frac{\left(\sum_{i \in I_R} g_i\right)^2}{\sum_{i \in I_R} h_i + \lambda} - \frac{\left(\sum_{i \in I} g_i\right)^2}{\sum_{i \in I} h_i + \lambda}\right) - \gamma \tag{7}$$

where $I$, $I_L$ and $I_R$ represents instances in the current node, left child and right child, respectively. $\gamma$ is the cost of making split. Based on the calculated gain, the tree is split iteratively to optimize the objective function, continuing until a stopping criterion is reached. The final prediction across all the tree is given by:

$$\hat{y}_i = \hat{y}_i^{(0)} + \sum_{k=1}^{K} \eta f_k(x_i) \tag{8}$$

where $\hat{y}_i^{(0)}$ is the initial prediction and $\eta$ is the learning rate controlling the contribution of each tree.

## 2.3 XGBoost Hyperparameter

In machine learning, hyperparameter tuning is essential for managing model complexity and ensuring generalization to unseen data. The parameters colsample_bytree, colsample_bylevel, and colsample_bynode control feature sampling at different stages of tree construction, introducing randomness that reduces feature dependency and helps prevent overfitting. The subsample parameter specifies the fraction of training data used to build each tree, adding further randomness. In our training, all four parameters were set to 0.5 to enhance model robustness. We set max_depth to 0, effectively removing any depth limit. To control complexity, min_child_weight was set to 200, which restricts splits unless a node contains enough instances, thereby avoiding splits based on small or noisy data. The lambda parameter was set to 500 to apply strong L2 regularization, penalizing large weights and discouraging unnecessary splits. A learning_rate of 0.3 was used to scale each tree's contribution, balancing faster convergence with controlled overfitting. All hyperparameters were selected using a random search strategy, with definitions referenced from the official XGBoost documentation.

## 2.4 Area Accuracy ($A_a$)

Traditional error metrics are ill-suited for comparing actual soil profiles with profiles predicted duriing inversion analysis due to differences in dimensions. To address this issue, we used a custom accuracy metric, formulated as follows:

$$A_a = \left(1 - \frac{A_i}{A}\right) * 100\ \% \tag{9}$$

The variable $A_i$ denotes the internal area between the actual and predicted soil profiles up to a depth of 40 m, while $A$ is the area of the actual soil profile.

## 3.2 Performance Evaluation of XGBoost Model

During training, the dataset of 10 million synthetic profiles was split into training, validation, and test sets, with 5% reserved for testing and 5% of the remaining data used for validation. The model achieved an average coefficient of determination ($R^2$) of 0.9831 on the training set and 0.9786 on the validation set. It was trained for 200 boosting rounds, achieving a final RMSE of 35.43 on the training set and 39.90 on the validation set (Fig. 4). The similarity between the training and validation loss curves indicates minimal overfitting. The distribution of feature importance across frequencies from 1 Hz to 100 Hz for XGBoost is presented in Fig. 5. The feature importance is measured by weight, reflecting the number of times a feature is used to split the data across all trees. The importance decreases steadily as frequency increases, with a slight rise near the higher end. This pattern suggests that the model assigns higher importance to lower frequencies while still considering higher frequencies to a lesser extent. The absence of sharp spikes indicates that no single frequency dominates, implying that the model effectively utilizes a broad frequency range for its predictions. The predictions for each layer on the test dataset are presented in Fig. 6, where the mean absolute percentage error (MAPE) is used as the evaluation metric. In each subfigure, the x-axis represents the true shear wave velocity, while the y-axis shows the predicted values. The red line denotes perfect predictions where true and predicted values align. Although MAPE is calculated on the entire test dataset, only 5,000 soil samples are plotted (Fig. 6) for clarity. The maximum MAPE of 6.53% observed for the surface layer, while the minimum MAPE of 0.47% was achieved for the half-space. Use of low frequencies enables our XGBoost Model to better capture intricacies of the deeper layers, resulting lower MAPE for the half-space. This enhances sensitivity to half-space, which leads to more accurate predictions.

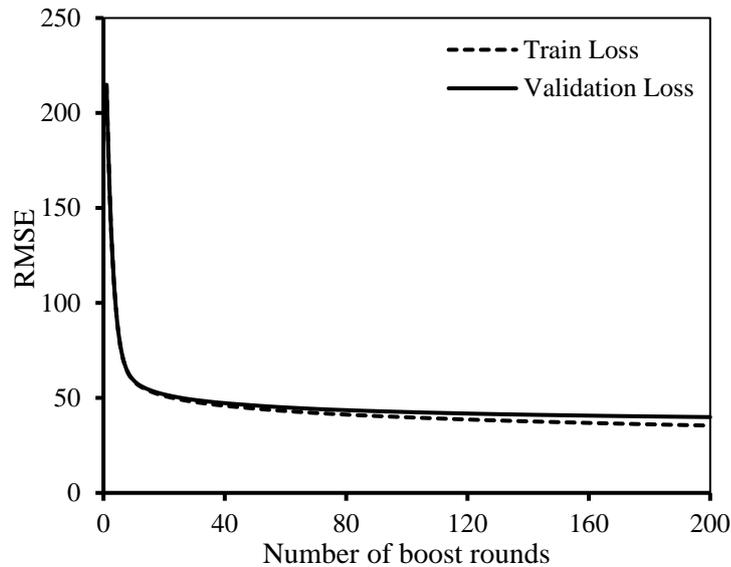

**Fig. 4.** Root Mean Square Error (RMSE) versus number of boosting rounds for the XGBoost model, exhibiting convergence of training and validation loss.

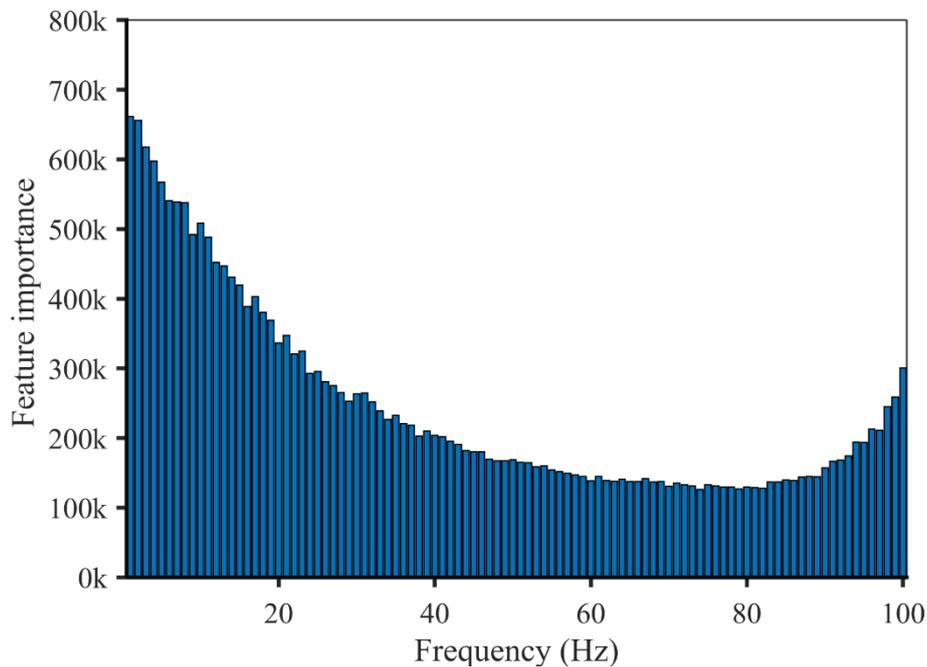

**Fig. 5.** Distribution of feature importance across frequencies (1–100 Hz, in 1 Hz increments), demonstrating how the model prioritizes different frequencies, with higher values indicating greater importance.

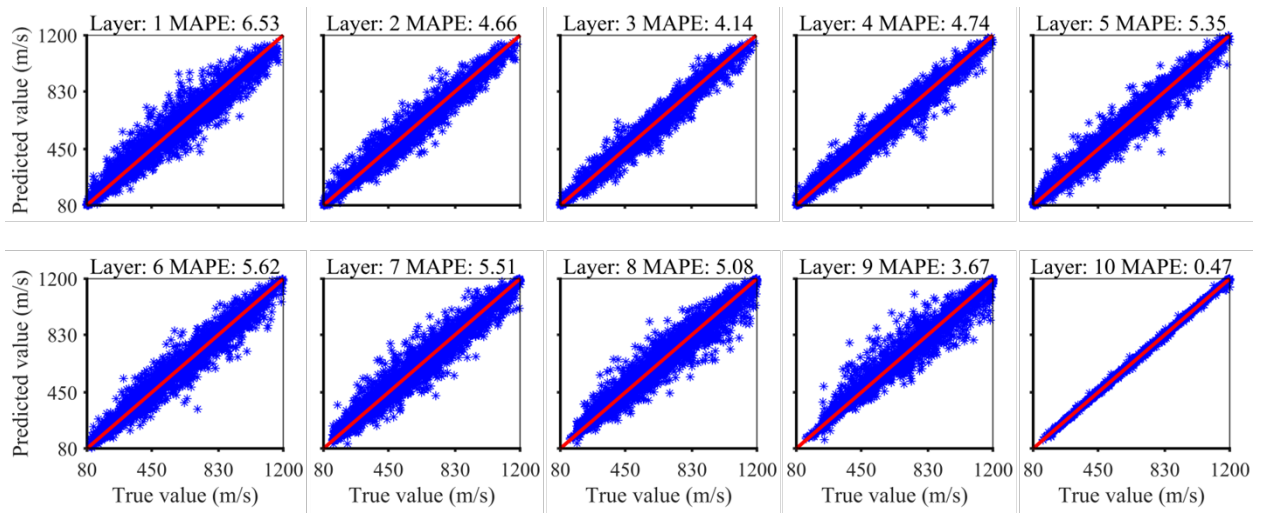

Fig. 6. Predicted shear wave velocity for each transformed soil layer in the test dataset. Lower Mean Absolute Percentage Error (MAPE) values indicate better accuracy.

Note that, each layer's MAPE does not indicate how well the XGBoost model captures the overall soil profile. To assess this, we used the area accuracy metric on the test dataset. The distribution of area accuracy value (Fig. 7) indicates that 99.93% of soil profiles achieved accuracy values above 80%. Additionally, 99.45%, 96.14%, and 77.85% of profiles exceed thresholds $A_a$ values of 85%, 90%, and 95%, respectively. Therefore, the proposed XGBoost model effectively recovers complete soil profiles with high precision. Fig. 8 illustrates the dispersion curve analysis, where the root mean square percentage error (RMSPE) quantifies the curve-fitting error between the measured and recovered dispersion curves. The recovered curves are obtained using forward modelling based on the XGBoost predicted shear wave velocities. Notably, 48.27% of soil profiles achieved an RMSPE below 1, while 85.25% have an RMSPE below 5, demonstrating the model's strong performance across diverse scenarios.

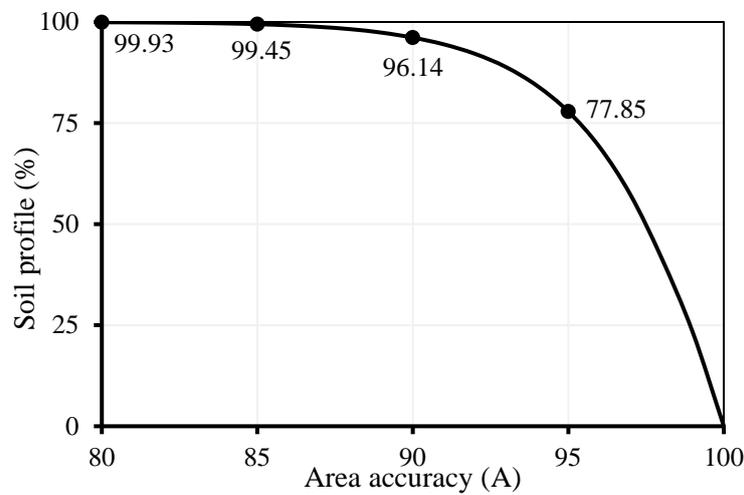

**Fig. 7.** Area accuracy analysis for the test dataset: the percentage of soil profiles exceeding specified thresholds. Higher values indicate better model performance.

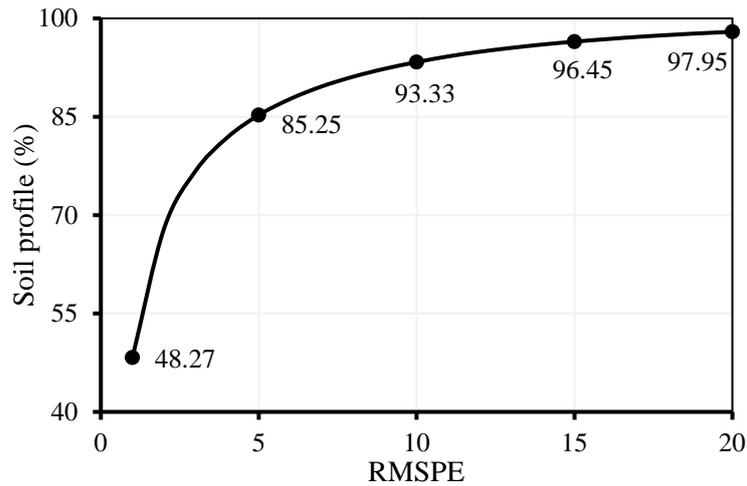

**Fig. 8.** Dispersion curve analysis for the test dataset: the percentage of soil profiles with Root Mean Square Percentage Error (RMSPE) below specified thresholds. Lower values indicate better model performance.

## 3 RESULTS

The use of 10 million diverse training profiles, along with the standardization of the soil model, enables our XGBoost framework to accurately predict shear wave velocity and layer thickness for a diverse range of subsurface strata. The accuracy, effectiveness, and generalization capability of the proposed XGBoost inversion tool are demonstrated using 14 soil profiles adopted from well-established studies published by renowned authors. Altogether, these profiles represent a wide range of testing scenarios, including both regularly and irregularly dispersive cases. Unlike existing machine learning methods, which perform well only over a narrow range of profiles, this broad set of profiles is adopted to demonstrate the proposed model's robustness and applicability in general testing scenarios. The inversion results are compared against widely used commercial software and the state-of-the-art global search-based tool, Geopsy. To ensure the half-space depth is close to 30 m, the parameter depth conversion ratio for LSM has been adjusted accordingly. On the other hand, GSM relies on a predefined search space within which it looks for an optimal solution. We adopt the layering ratio approach (Cox & Teague, 2016) to define the optimal search space, performing multiple inversions with varying layering ratios (1.2, 1.5, 2, 3, 3.5, and 5) due to the unknown number of soil layers. A depth constraint of 30 m is imposed for the half-space, with shear-wave velocity bounds set between 0.5 times the minimum and 1.5 times the maximum phase velocity. To ensure robustness and stochastic variability, three independent trials with random seeds were performed for each layering ratio. Note that dispersion curves are highly sensitive to near-surface layers, and small differences in these layers can cause significant misfits. Therefore, a large dispersion curve misfit alone cannot confirm that the profile is far from the true one.

**Profile I:** Two Layer Model

The profile I is adopted from Xu et al., (2007), represents the simplest regularly dispersive soil profile consisting of a single layer overlying a half-space (Fig. 9). It serves as an illustrative example to demonstrate how the proposed standardization technique predicts a 2-layer profile with reasonable accuracy using a 10-layer model by subdividing the original layers into thinner sublayers with equivalent properties. For GSM, using a higher layering ratio reduces the number of soil layers, thereby shrinking the search space dimensionality. Since each inversion trial performs 50,000 forward model evaluations, a reduced dimensionality results in greater sampling density within the defined search space. This increased sampling density enhances the likelihood of identifying the true soil profile, turning it into a well-constrained problem, which explains the GSM's performance that closely matches the true profile. The local search method achieves the least area accuracy as it fails to accurately capture the transition zone characteristics. In XGBoost, the target dispersion curve is used as input to determine the corresponding leaf nodes in each tree, and the final prediction is obtained by summing the weights of these leaf nodes across all trees in the ensemble. The XGBoost does not explicitly optimize predictions to match dispersion curves through forward modeling. As a result, the predicted soil profiles may not perfectly replicate the input dispersion curves (Fig 9a), but they often remain close to the true profiles due to the model's ability to learn stable, data-driven relationships. Although the XGBoost model was trained to output only 10-layer profiles, using the proposed soil standardization technique, it successfully adjusted the shear wave velocities across layers to predict this two-layer profile with reasonable accuracy. It clearly outperforms the LSM approach, and the predicted soil model is on par with GSM's best-fit model. Furthermore, compared to LSM, it demonstrates better capability in representing the half-space. Thus the proposed framework found to perform well in the most extreme senarios.

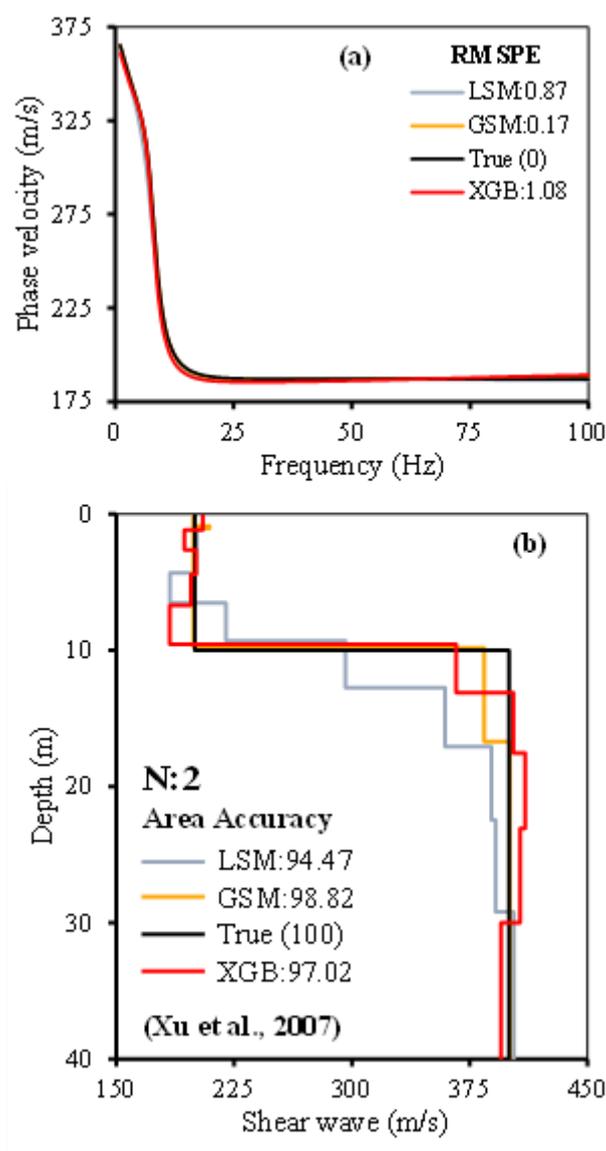

**Fig. 9.** Inversion results for Profile I, (a) recovered dispersion curves, and (b) predicted shear wave velocity profiles

**Profile II:** Multiple Layers Regularly Dispersive Model

Profile II is adopted from Passeri et al., (2021), represents a regularly dispersive soil profile with the half-space located at a depth of 13.9 m . It is characterized by a low shear wave velocity in the surface layer and gradually increasing to a higher velocity in the underlying half-space. While the previous example evaluated the proposed model on a 2-layer profile, this example examines the other extreme by testing the proposed framework's ability to accurately predict a complex 9-layer profile. Moreover, the fundamental mode dispersion curve exhibits complex silhouettes (Fig. 10), challenging to match even for a global search method. The LSM method exhibited a significantly large dispersion misfit, and the recovered curve deviated noticeably from the target dispersion curve. In contrast, the GSM and the proposed XGBoost matched the target dispersion curve more accurately. Note that, although statistically the XGBoost displayed a slightly higher RMSE of 3.07, it was able to capture the overall shape of the target dispersion curve and was visually almost indistinguishable. The LSM approach converges to a suboptimal solution, wrongly predicting the half-space depth, which contributed to its lower area accuracy. On the other hand, the GSM and the proposed XGBoost performed extremely well and attained an impressive area accuracy of above 96. Note that we trained our model using a 2 to 6 layer profile converted into a ten-layer profile; therefore, technically the XGBoost has never seen a nine-layer model, yet it performs extremely well, validating the effectiveness of the proposed soil profile standardization technique.

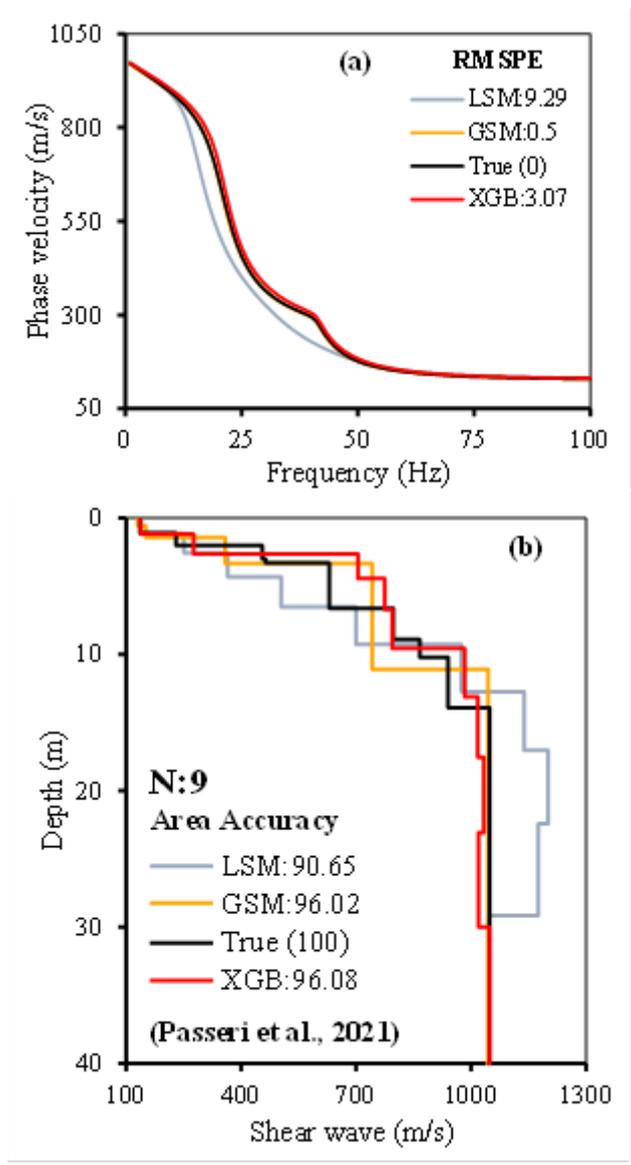

**Fig. 10.** Inversion results for Profile II, (a) recovered dispersion curves, and (b) predicted shear wave velocity profiles.

**Profile III:** Low Velocity Layer at Shallow Depth

Profile III (Li et al., 2024) represents a four-layer irregularly dispersive soil model, characterized by a low-velocity layer located at a shallow depth of 5 m. A distinctive feature of irregularly dispersive profiles is the presence of a hump in the dispersion curve, as observed in Fig. 11. Such complex soil profiles are always challenging to predict accurately by either local or global inversion schemes. The LSM method fails to accurately recover the true soil profile, even though it fits the dispersion curve well. The LSM does not search the solution space thoroughly and depends heavily on the starting model, which may lead it to a nearby local minimum. On the other hand, GSM captures the low velocity layer effectively; however, it fails to locate the layer above the half-space, leading to a slightly lower area accuracy compared to the proposed XGBoost model. The XGBoost learns patterns directly from the data without using forward modeling to minimize the dispersion curve error, and in this case, it is able to capture the overall trend effectively. As our method assumes that each layer has a constant thickness, the area accuracy can never theoretically reach 100% unless the true profile has the same number of layers with matching thicknesses. Despite this limitation, the method still performs better than state-of-the-art GSM.

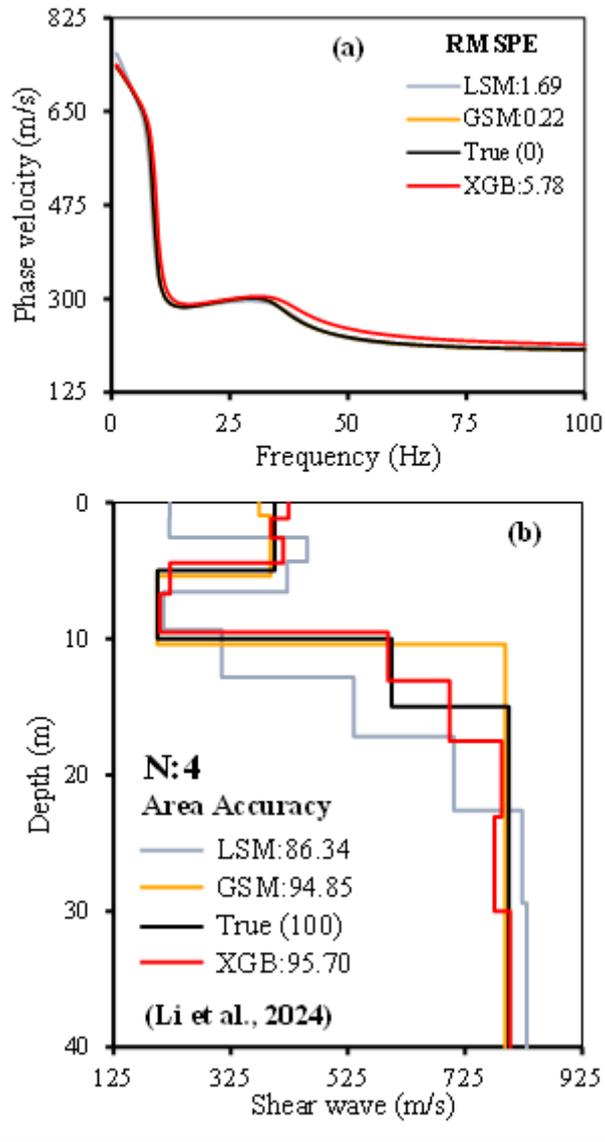

Fig. 11. Inversion results for Profile III, (a) recovered dispersion curves, and (b) predicted shear wave velocity profiles

**Profile IV:** Low Velocity Layer at Deeper Depth

Profile IV represents a four-layer irregular dispersive soil profile obtained from Wang et al., (2022), characterized by a low-velocity layer (LVL) located at 10 m depth with a thickness of

5 m. The shear wave velocity in the first two layers is 200 m/s and 600 m/s, before decreasing to 400 m/s for LVL. It is doubling from LVL to half-space, jumping from 400 m/s to 800 m/s. This significant velocity jump poses challenges to any inversion approach currently in use. LSM produces a reasonably good fit with minor deviations in matching the shear wave velocity, especially for half space (Fig 12). In contrast, the GSM approach exhibits a larger deviation from the true profile as it tends to smooth out the velocity variations. This smoothing resulting in a less accurate recovery of the Shear wave velocity profile as iIt underestimate the velocity of second layer and over estimalte for the LVL layer. Note that, we have manually implimented special constraint to prohibit thin layer generation for GSM typically observed in case of irregularly dispersive profile. Otherwise, the GSM results would have been further effected by them. On the other hand, the proposed XGBoost provides the closest match to the true profile with a area accuracy score of 92.36, outperforming both LSM (90.30) and GSM (80.55). It able to capture the all velocity revearsal with greater accuracy. During ML model training, our proposed constraint inherently prohibits generation of thin unrealistic profiles, which reduces the chance of artifacts common to GSM approach. Overall, the results highlight the superior predictive ability of the proposed XGB framework in resolving sharp velocity variations and its ability to capture key features without rigorus inversion analysis. Therefore, the present work hold great potential is saving significan computational time and cost.

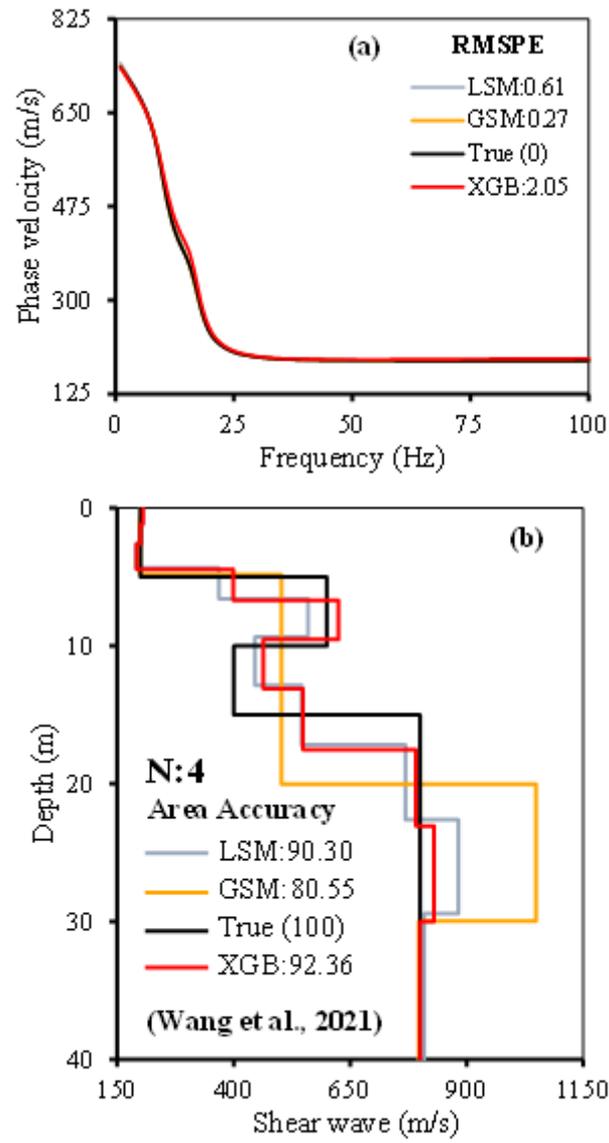

**Fig. 12.** Inversion results for Profile IV, (a) recovered dispersion curves, and (b) predicted shear wave velocity profiles.

### 3.4 Inversion Results for Additional Profiles:

Often, the existing ML research paper apply their method to a few contrived soil profiles to hide their limited general applicability. Either number of predicted soil layers will be fixed, or the predicted model will have a narrow velocity range close to their training model. Our novel soil standardization technique and 10 million training profiles enable the proposed XGBoost model to be applied to a wide range of soil profiles with varying layer numbers and shear wave velocities. The consistent performance of XGBoost across diverse soil conditions is further illustrated using 10 additional soil profiles with their inversion results presented in Fig. 13. To eliminate any sort of bias, we have adopted these profiles only from the literature published by the reputed authors. Each profile depicts a distinct soil configuration and poses unique challenges for inversion. Together, these 14 (4+10) profiles are comprising of 2 to 10 layers with shear wave velocities ranging from 103 m/s to 1048 m/s, layer thicknesses varying from

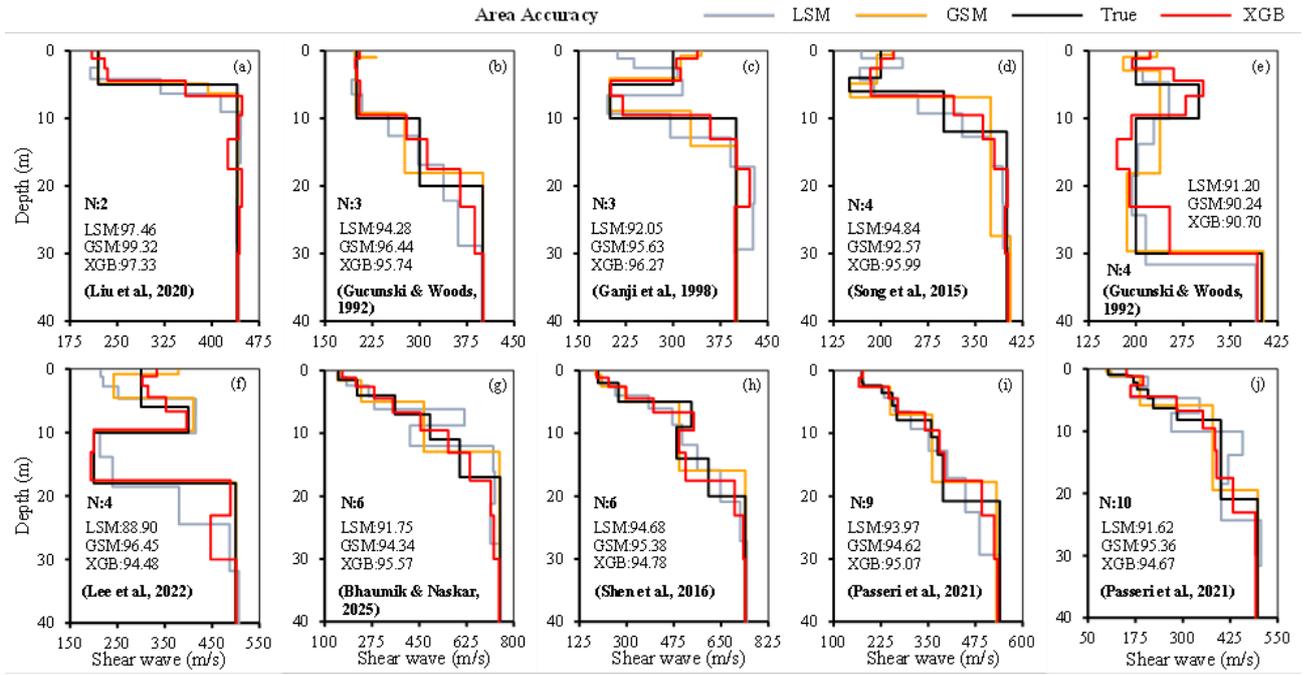

**Fig. 13.** Comparison of inverted profile predicted by local search, global search, and XGBoost methods for (a) Profile V, (b) Profile VI, (c) Profile VII, (d) Profile VIII, (e) Profile IX, (f) Profile X, (g) Profile XI, (h) Profile XII, (i) Profile XIII, and (j) Profile XIV.

0.2 m to 20 m. The half-space depth varies from 5 m to 30 m. The consecutive inter-layer velocity contrasts vary between -200 m/s to 400 m/s. In percentage terms, these consecutive layer velocity changes will result in -50% and +200% (Table. 1).

**Table 1.** Soil parameter range for 14 published profiles.

| Soil Parameter | Minimum | Maximum |
| --- | --- | --- |
| Number of layers | 2 | 10 |
| Shear wave velocity (m/s) | 103 | 1048 |
| Layer thickness (m) | 0.2 | 20 |
| Velocity difference (m/s) | -200 | 400 |
| Velocity difference (%) | -50 | 200 |
| Half-space depth (m) | 5 | 30 |

In the current state of practice, GSM is considered as a superior option compared to LSM. The overall results support this, as GSM outperform LSM in majority of cases. However, the proposed XGBoost model outperforms the LSM and produce similar result to those of Geopsy. For all 14 diverses cases, either the proposed method perfoed best or its performance is within a close range of the best-performing inversion tools. This is a commendable achievement, considering that existing machine learning methods have so far demonstrated good performance only over a limited range of soil profiles close to their training model. We would like to remind readers; the model is trained only using synthetic profile containing 2-6 layers that are converted into 10-layer format. Therefore, some of the profiles it accurately predicted are significantly different and were never seen during the training period. Despite this, the proposed XGBoost outperforms the state-of-the-art tool Geopsy, indicating its general applicability.

**Computation Time**

The local search method uses the Jacobian matrix to update the initial soil profile, which reduces the dispersion curve misfit in each iteration. For this reason, it converges faster and requires fewer forward model evaluations. In contrast, the global search method explores the entire predefined parameter space to avoid suboptimal solutions. While existing LSM-based approaches complete the inversion in under a minute, the improved accuracy of GSM methods comes at a much higher computational cost, typically around an hour. In contrast, our trained XGBoost model is extremely fast, requiring only a fraction of a second while delivering performance that matches or surpasses state-of-the-art GSM tool. The XGBoost model directly maps the input dispersion curve to the soil profile without forward computation, making it several orders of magnitude faster than the local search method (Table. 2). The average computation times for the local search method, global search method, and XGBoost corresponding to 14 selected soil profiles are 9.69, 5850, and 0.008 seconds, respectively. Thus, the proposed XGBoost reduces the computational time by several orders of magnitude. This will be especially beneficial in processing a large set of survey data, as it has the potential to save data processing time from a few weeks to a day. All models were run on a desktop PC with an 11th Gen Intel Core i7-11700K CPU (8 cores, 16 threads, 3.60 GHz) and 64 GB of DDR4 RAM (3200 MHz), without GPU acceleration.

**Table 2.** Comparison of computational time for different inversion methods.

| Profile | LSM (s) | GSM (s) | XGBoost (s) |
|---|---|---|---|
| Profile I | 2.2 | 3544 | 0.008 |
| Profile II | 35.3 | 12571 | 0.007 |
| Profile III | 9.7 | 2926 | 0.007 |
| Profile IV | 12.1 | 4045 | 0.007 |
| Profile V | 2.2 | 6256 | 0.014 |
| Profile VI | 2.1 | 3481 | 0.008 |
| Profile VII | 6.9 | 3273 | 0.008 |
| Profile VIII | 3.7 | 4444 | 0.007 |
| Profile IX | 2 | 7330 | 0.007 |
| Profile X | 11.8 | 4540 | 0.007 |
| Profile XI | 20 | 6641 | 0.007 |
| Profile XII | 3.9 | 5548 | 0.007 |
| Profile XIII | 2.6 | 6361 | 0.007 |
| Profile XIV | 21.1 | 10938 | 0.005 |

**Conclusion**

The study addresses major limitations of existing machine learning based surface wave inversion methods, which are often applicable to a fixed number of layers and a narrow shear-wave velocity range. The paper presents a novel soil profile standardization technique to generate training data for an ML model and proposes a regression-based XGBoost framework to predict S-wave velocity and layer thickness. By converting variable-layer profiles into a uniform ten-layer format with geometrically increasing thicknesses, the proposed method ensures consistent model output, thus avoiding explicit layer number estimation. Furthermore, we proposed implementing constraints during training generation to prohibit abrupt velocity changes and prevent the generation of unrealistic thin layers typically observed for GSM approach. Thus, it can maintain high prediction accuracy across diverse soil conditions. The efficiency of the proposed model is demonstrated using 14 soil profiles adopted from published literature and by comparing the prediction result with LSM and state-of-the-art GSM tool Geopsy. They contain both regular and irregularly dispersive profiles with high and low velocity strata present in complex multi-layered structures. Together, these profiles represent a wide variation in the layer numbers, thicknesses, and shear wave velocities. The present model performs accurately and captures shear-wave velocity reasonably well, even for the most difficult profiles. The proposed model successfully predicts complex profiles that are significantly different from the training data, demonstrating its general applicability to diverse conditions. Overall, it outperforms the current state-of-the-art GSM tool Geopsy in the 50% cases. This impressive accuracy comes with added computational benefit as the proposed model only takes fraction of the second. Furthermore, although at present we have trained the model for 1hz to 100 Hz frequency band, the proposed model can be trained for any frequency band as per the requirements. Thus, the current study has huge potential, as it can be employed to process large datasets like 2D/3D surface wave surveys and DAS based surveys.

**Acknowledgements**

**Figure Caption**

**Fig. 1.** Flowchart of the training data generation process.

**Fig. 2.** The soil profile standarization technique: Transformation of a four-layer soil profile into a standardized ten-layer format..

**Fig. 3.** Flowchart illustrating the training and prediction workflow of the XGBoost model.

**Fig. 4.** Root Mean Square Error (RMSE) versus number of boosting rounds for the XGBoost model, exhibiting convergence of training and validation loss.

**Fig. 5.** Distribution of feature importance across frequencies (1–100 Hz, in 1 Hz increments), demonstrating how the model prioritizes different frequencies, with higher values indicating greater importance.

**Fig. 6.** Predicted shear wave velocity for each transformed soil layer in the test dataset. Lower Mean Absolute Percentage Error (MAPE) values indicate better accuracy.

**Fig. 7.** Area accuracy analysis for the test dataset: the percentage of soil profiles exceeding specified thresholds. Higher values indicate better model performance.

**Fig. 8.** Dispersion curve analysis for the test dataset: the percentage of soil profiles with Root Mean Square Percentage Error (RMSPE) below specified thresholds. Lower values indicate better model performance.

**Fig. 9.** Inversion results for Profile I, (a) recovered dispersion curves, and (b) predicted shear wave velocity profiles.

**Fig. 10.** Inversion results for Profile II, (a) recovered dispersion curves, and (b) predicted shear wave velocity profiles.

**Fig. 11.** Inversion results for Profile III, (a) recovered dispersion curves, and (b) predicted shear wave velocity profiles.

**Fig. 12.** Inversion results for Profile IV, (a) recovered dispersion curves, and (b) predicted shear wave velocity profiles.

**Fig. 13.** Comparison of inverted profile predicted by local search, global search, and XGBoost methods for (a) Profile V, (b) Profile VI, (c) Profile VII, (d) Profile VIII, (e) Profile IX, (f) Profile X, (g) Profile XI, (h) Profile XII, (i) Profile XIII, and (j) Profile XIV.

**Table Caption**

**Table 1.** Soil parameter range for 14 published profiles.

**Table 2.** Comparison of computational time for different inversion methods.